\documentclass[]{spie}
\usepackage[]{graphicx}
\title{The ALMA archive\\and its place in the astronomy of the future}

\author{Felix Stoehr\supit{a}, Mark Lacy\supit{b}, St\'{e}phane Leon\supit{c}, Erik Muller\supit{d}, Alisdair Manning\supit{a}, Christophe Moins\supit{a}, Dustin Jenkins\supit{e}
\skiplinehalf
\supit{a}European Southern Observatory, Karl-Schwarzschild-Str 2, 85748 Garching, Germany \\
\supit{b}National Radio Astronomy Observatory, 520 Edgemont Road, Charlottesville, VA 22903-2475, USA\\
\supit{c}Joint ALMA Observatory, Alonso de C\'{o}rdova 3107, Vitacura 763 0355, Santiago, Chile\\
\supit{d}National Astronomical Observatory of Japan, 2-21-1 Osawa, Mitaka, Tokyo 181-8588, Japan\\
\supit{e}Canadian Astronomy Data Center, 5071 West Saanich Road, Victoria, V9E 2E7, Canada\\
}

\begin{document}
\maketitle

\begin{abstract}
The Atacama Large Millimeter/submillimeter Array (ALMA), an international partnership of Europe, North America and East Asia in cooperation with the Republic of Chile, is the largest astronomical project in existence. While ALMA's capabilities are ramping up, Early Science observations have started.

The ALMA Archive is at the center of the operations of the telescope array and is designed to manage the 200 TB of data that will be taken each year, once the observatory is in full operations. We briefly describe design principles.

The second part of this paper focuses on how astronomy is likely to evolve as the amount and complexity of data taken grows. We argue that in the
future observatories will compete for astronomers to work with their data, that observatories will have to reorient themselves from
providing good data only to providing an excellent end-to-end user-experience with all its implications, that science-grade
data-reduction pipelines will become an integral part of the design of a new observatory or instrument and that all this evolution will have a
deep impact on how astronomers will do science. We show how ALMA's design principles are in line with this paradigm.
\end{abstract}


\section{INTRODUCTION}
ALMA is a radio interferometer with 66 antennas placed at 5000m elevation in the Atacama Desert in northern Chile. In full operations it will allow baselines of up to 16km corresponding to a spatial resolution of 6 mas at 675GHz. ALMA will produce data at an average rate of about 200TB/year with short-term peak data-rates of ten times as much. Currently ALMA Early Science operations are in progress and  observations of the third proposal cycle are about to begin.

\begin{figure}
   \begin{center}
   \begin{tabular}{c}
   \includegraphics[width=12cm]{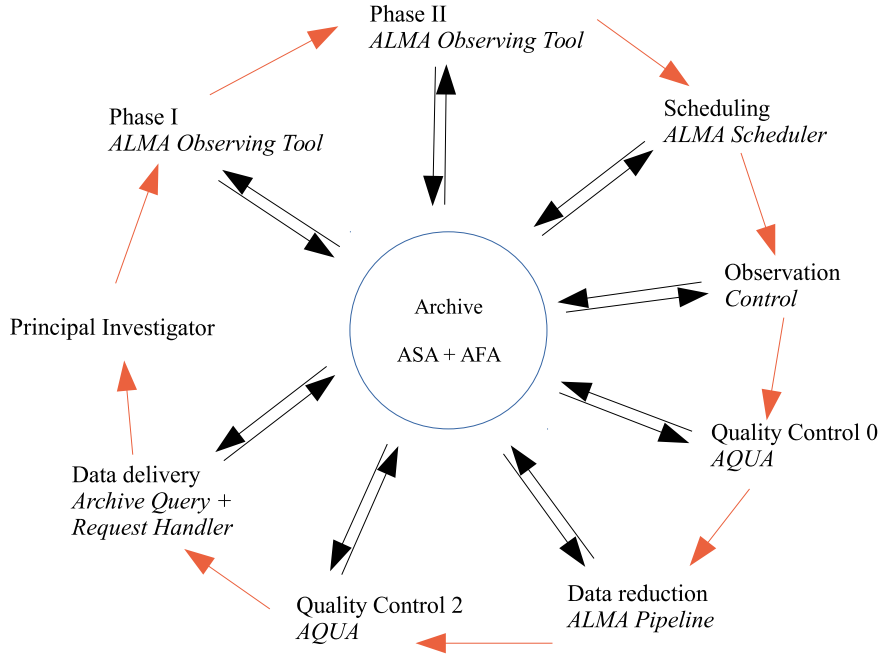}
   \end{tabular}
   \end{center}
   \caption[example]
   {The ALMA archive is in the centre of the ALMA data flow.\label{fig:dataflow}}
   \end{figure}

The ALMA archive is at the centre of the ALMA data flow (Fig. \ref{fig:dataflow}). It is a combined database and binary data storage system that is accessed by the different software subsystems through the same software layer. The ALMA archive is divided into two parts. The ALMA Frontend archive (AFA), which provides the core persistence functionality and into the ALMA Science Archive (ASA). The latter holds a tiny subset of metadata of the AFA and provides access to external interfaces like the Archive Query interface and Virtual Observatory (VO) tools.

The ALMA archive is a fully distributed database and data storage-system, with operational parts at the Operations Support Facility (OSF), the Santiago Central office (SCO) and the three ALMA Regional Centres (ARCs). The main archive is situated at the SCO, holding all science data as well as commissioning and operations data including the monitoring data of the array itself. Full copies of the main archive are stored at the three ARCs. From the archives at the ARCs data is distributed to PIs and archival researchers. They also serve as off-site backup copies of the main archive, allow ALMA staff at the ARCs to give user support, to do quality control (QA3) and to help improving the ALMA pipeline.

In contrast to the original design where the data transfer between SCO and the ARCs was supposed to happen via shipped hard-disks, bandwidth costs have dropped enough in the meantime to already now allow all ALMA data to be transferred between the ALMA sites over the network in near-real time. For more details about the ALMA archive and ALMA software, please refer to Espada et al.\cite{DanielEspada}.

In the following we describe, the rationale, design principles, policies and status of the ALMA Science Archive and then discuss the evolution of astronomy as a whole as well as how ALMA fits into that picture.

\section{SCIENCE ARCHIVE}
\label{sec:asa}
\subsection{RATIONALE}
\label{sed:rationale}
The success of ALMA and of any other astronomical facility is measured by the scientific output of the community. As thus by construction the facility can not assure their own success directly, the way to improve success is to render the end-to-end user-experience for PIs and archival researchers as perfect as possible. This does of course include the quality of the data, but goes far beyond as we will argue in section \ref{sec:astronomy}.

Creating a Science Archive that allows archival researchers to easily discover and retrieve the data they can use is part of a good end-to-end user-experience and thus helps to maximize the scientific return of the facility. In addition, we are convinced that creating or improving the Science Archive of a facility is a highly cost effective way of augmenting the scientific impact. As the data have to be provided for the PI anyway, the additional cost for a good Science Archive is a tiny fraction of the total operations cost. Depending on the facility perhaps of the order of 1\%. The scientific impact on the other hand can be huge. The most prominent example being the HST archive where the total number of archival papers even outnumbers the PI papers\footnote{https://archive.stsci.edu/hst/bibliography/pubstat.html}.

\subsection{DESIGN PRINCIPLES}
The most important design principle of the ALMA Science Archive is to make sure that the archive, in particular the query interface, speaks the language of the scientists. It must allow to query for physical concepts in the categories Position, Energy, Time and Polarization. I.e. the four properties a photon can carry. Orbital angular momentum being neglected here.

Querying by physical concepts in particular means to favour spatial resolution over antenna configuration, frequency over receiver-band and spectral resolution over spectral mode. Once the holdings are fully described with physical concepts, it is also easy to provide additional services like VO searches or cutout services.

As much as possible, the ASA should only provide the relevant information at any given time. For example, help text for a query field is only provided when the user is about to enter a value into that field. Even the input fields themselves are hidden as long as they are not needed. Also, per default only the publicly available science observations are presented and only the most relevant columns in the results view are shown.

Wherever possible hurdles should be removed. User-registration should be required only when proprietary data is to be downloaded and in those cases, the interface should send the user to the login page automatically. Installation of additional software should be avoided.

Search is a two step process which often is forgotten. After the search is executed and the result list is obtained, users need to be able to quickly decide which of the data returned are relevant for them. For this to be possible, users need to be able to drill down into the search results and be able to manipulate them easily. Next to previews, which are the single most important help for users, they should be able to sort and sub-filter, to select additional more specialized columns and to obtain additional information for cells of the result grid, i.e. more detailed spectral information or the abstract text of the proposal.

Showing the users an updated view of the parameter space that remains after they have entered their constraints is clearly extremely helpful. Commercial sites do this by using facets. The complication for astronomical searches comes from the fact that the useful facets are all values and not just keywords. In that sense, too, astronomical searches differ widely from the problem search engines like Google are solving.

The main principles for the database content are completeness, correctness, consistency and homogeneity. Search columns must be completely filled for all rows, NULLs are not acceptable. Values given have to be correct, even a small number of false positives can ruin the user-experience. Referential integrity is crucial to any database, and is essential for building services on top.

Although classical user-interfaces will stay relevant for the near future, full programmatic access to both, metadata and data is indispensable. Running fully automatic query, access and analysis scripts will be a standard use-case in the future. Programmatic access can be realized through VO services for example.

For the ASA, we chose to create a separate set of optimized database tables which contain a tiny but curated amount of the total metadata available.

\subsection{POLICIES}
A careful choice of policies supports the design of the the Science Archive. In order to reduce hurdles, ALMA indeed selected to allow anonymous access to all public data and metadata. Furthermore there will not be any private metadata in the ASA. Users are able, for example, to see metadata of the observations that are still in the proprietary period. PIs are not treated differently from archival researchers. In particular they will use the same tools and get both the exact same data. The difference reduces to PIs having earlier data access. In order to allow to track publications and to create the link between data and publications (see Grothkopf et al.\cite{Grothkopf} in this volume) authors are required to place the official ALMA statement into the acknowledgement section of their publications. This statement contains the ALMA data tag in ADS format\footnote{http://vo.ads.harvard.edu/dv/}. Readers can copy\&paste this tag into the ASA interface and obtain the corresponding data. In the future, archival researchers browsing the data, will get links to the corresponding publications. Finally, simplicity is gained by deciding that metadata will be ingested into the ASA as soon as the first data of a project has been observed.

Users should never be tempted to give their ALMA Science Portal (SP) passwords away. To this end ALMA has implemented a project delegation feature which allow PIs to give access to an other registered ALMA user. In the future, delegation will also be possible for triggering of Target Of Opportunity Proposals.

\subsection{BUILDING BLOCKS}
In addition to the database scheme, the ASA consists of 4 building blocks. The Harvester extracts a small subset of the metadata from the AFA and writes a into the ASA database in a homogenized and curated way. This software is written in Java and runs twice a day. The Archive Query (AQ)\footnote{http://almascience.org/aq} interface, which is part of the SP, is built using Java Server Pages (JSP) and JavaScript and provides a classical form-based search interface to selected columns of the ASA database. This package is currently mainly developed at the Canadian Astronomy Data Center (CADC) in Victoria. Once the user hits the search button, the query is transformed into the Astronomy Data Query Language (ADQL) a dialect of SQL and passed to the Query Backend. Developed at CADC, too, the OpenCADC Tap software is a generic Java code allowing to query Relational Databases and return results in VO format. The result of this query is then passed back to the query interface for display. Thus, internally, all ASA queries are VO queries which also means that all user-services querying the archive will use the same query backend. This reduces cost, simplifies the the set-up, allows for easy maintenance and uses existing standards. Once the user has selected data and, if necessary has been sent to the authentication page of the SP, she or he will be sent to the fourth pillar of the ASA, the ALMA Request Handler (see Hill et al.\cite{2009ASPC..411..410H}) (RH). The RH manages user requests for the download of data or for the processing of data. It stores the users' requests under the user's account and provides the access control on a file level and allows for different download methods. The standard method is a download manager (Java applet) which allows the user to download several files in parallel in order to optimize the use of the available bandwidth. Paused or broken downloads can be resumed. It also offers a download script which is meant for users who want to download the ALMA data directly onto a processing environment without having to use a web browser. The third download method will be offered soon: users can request ALMA data to be shipped to them on USB hard drives in case their Internet connection is not good enough. The RH logs data access to be able to generate download statistics.

\subsection{STATUS}
\label{sec:status}
A first version of the AQ is online since Q1 2013. It features searches along the Position, Energy, Time and Polarization axes as well as some observation related parameters like PI name or project title. The search capabilities are still limited and will be improved in the future. The largest limitation comes from the fact that currently only raw-data can be searched for and, by construction, many result rows can appear for the same object. Since Q1 2014 downloads can happen through the RH. The products are contained in tar files which transport not only the data but also a directory tree following the ALMA data structure.

Once the ALMA pipeline is running in full production mode, science products can be ingested into the ASA directly and the search can be based on products instead of raw-data. This is also the moment when ALMA can offer previews on the AQ and when users will be able to access individual data cubes directly.

Further developments will bring improvements to the user interfaces, cutouts, the VO services SIAPv2, ObsTAP and TAP as well as the integration of externally developed software tools. The first such tool will be  like a powerful server-side visualization tool (PI: Erik Rosolowsky) which allows users to browse and manipulate the very large ALMA data cubes without having to download them to disk first. The second of such tools is a  a post-pipeline science-analysis tool (PI: Lee Mundy). This tool will run at JAO directly after the ALMA pipeline and will do source extraction, line-finding and science analysis.

   \begin{figure}
   \begin{center}
   \begin{tabular}{c}
   \includegraphics[width=12cm]{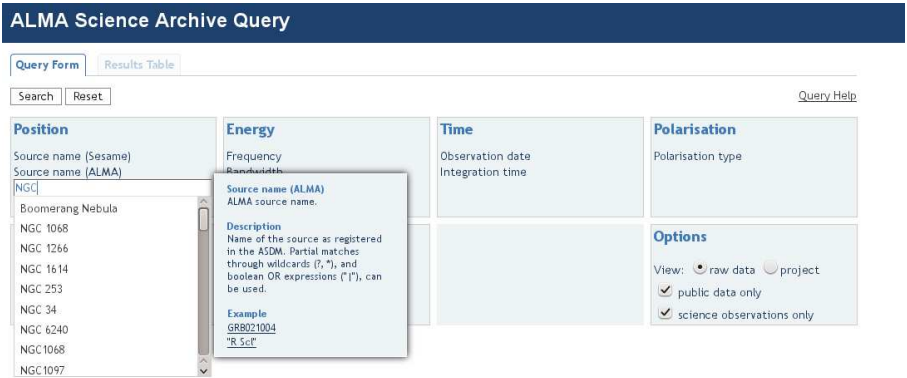}
   \end{tabular}
   \end{center}
   \caption[example]
   {Archive Query interface\label{fig:aq}}
   \end{figure}

   \begin{figure}
   \begin{center}
   \begin{tabular}{c}
   \includegraphics[width=12cm]{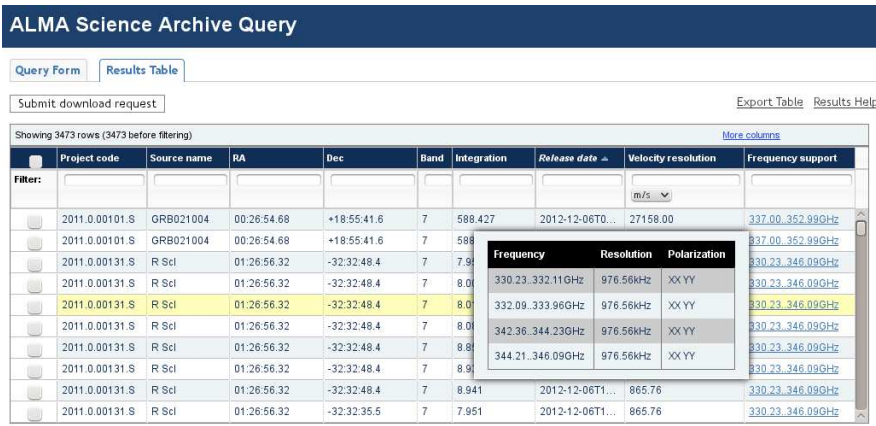}
   \end{tabular}
   \end{center}
   \caption[example]
   {ALMA Query interface results table\label{fig:res}}
   \end{figure}

   \begin{figure}
   \begin{center}
   \begin{tabular}{c}
   \includegraphics[width=12cm]{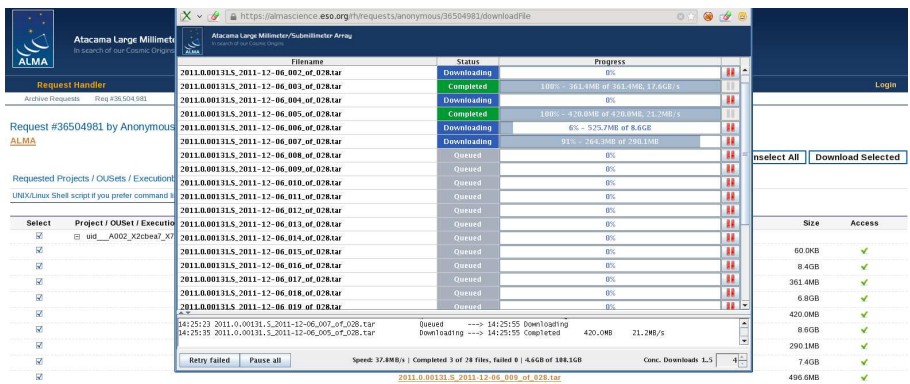}
   \end{tabular}
   \end{center}
   \caption[example]
   {ALMA Request Handler and Download Manager\label{fig:rh}}
   \end{figure}

\section{ASTRONOMY OF THE FUTURE}
\label{sec:astronomy}
\subsection{EVOLUTION OF OBSERVATORIES}
\begin{figure}[h]
   \begin{center}
   \begin{tabular}{c}
   \includegraphics[width=12cm]{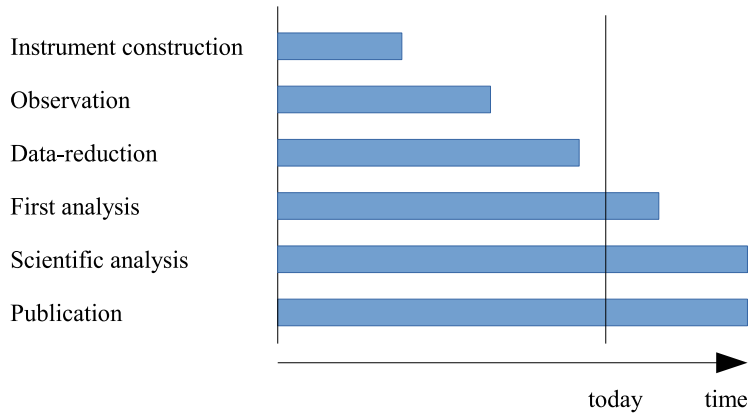}
   \end{tabular}
   \end{center}
   \caption[example]
   {Part of the workflow PIs typically do themselves -- past to future.\label{fig:evolutionworkflow}}
   \end{figure}

The way astronomy is been done has changed dramatically over time. Whereas in the beginning the whole process from building the instrument to the publication of the scientific results was carried out by the astronomers themselves, since then more and more of the workflow has been taken over by specialists (Fig. \ref{fig:evolutionworkflow} ). Driven by the increased technical complexity of observatories service-mode observing is standard since a few decades now.

Although there is no sharp transition, we argue here that even the times when astronomers did the data-reduction themselves has essentially ended and observatory data-reduction even for ground-based observatories will become the norm soon. Indeed, we are convinced that future instrument contracts will include a data-reduction pipeline as well as its maintenance for at least the life-time of the instrument as an integral part of the instrument design. Data-management costs will become a very significant fraction of the cost of an instrument. Ground-based facilities will follow the path laid out by spaced-based facilities like HST, XMM and Spitzer.

The next step of transferring work from the astronomers to the observatories, shifting part of the science analysis to the observatories, has already started. The Sloan Digital Sky Survey (SDSS) certainly made a huge step into this direction and is now followed by Gaia, PanStars and  LSST. But also observatories like XMM provide catalogs and fluxes for all their observations. As mentioned in section \ref{sec:status}, ALMA will have a science-analysis tool built into the standard data flow.

\subsection{EVOLUTION OF HARDWARE}
The key parameters of the data-flow hardware of telescopes, detector size, network bandwidth, memory, hard-disk capacity and CPU power are all growing exponentially following Moore-type laws. For telescopes which are generally planned to take in data at a linear scale, after an initial ramp-up, storage and processing costs drop again (see e.g. Schlegel\cite{2012arXiv1203.0591S}).

A closer look onto the exponential growth reveals, however, that the nature of the growth for Hard disks, Network speed (to some extent) and CPU speed is due to the increased total power, not the increased power of a single component. As a result, storage, network transfer and computing need to be parallelized in order to make use of the available growth.

Traditionally storage systems have been inherently parallel and the Next Generation Archive Storage (NGAS) used by ALMA is no exception. Data mirroring between the ALMA sites also has been set up using multiple parallel streams right from the beginning. However the data downloads to the users typically happened sequentially before the download managers were introduced in the CADC, ESO and ALMA Request Handlers. Similarly, although parallel computing is standard for decades now, astronomical data-reduction packages are generally not inherently parallelized. They rather rely on the fact that often astronomical data-reduction is embarrassingly parallel, i.e. that several reductions can be executed independently alongside. This method will come to an end as the user-waiting time for the reduction of a single dataset will exceed the acceptable time soon and we therefore predict that future astronomical data-reduction packages will be designed to be inherently parallel. Some tasks of ALMA's data-reduction software CASA \footnote{http://casa.nrao.edu} have already been parallelized.

\begin{table}[h]
\caption{Scalingrelations of typical hardware components}
\label{tab:fonts}
\begin{center}
\begin{tabular}{|l|l|} %% this creates two columns
%% |l|l| to left justify each column entry
%% |c|c| to center each column entry
%% use of \rule[]{}{} below opens up each row
\hline
\rule[-1ex]{0pt}{3.5ex}  & Moore-type law?  \\
\hline
\rule[-1ex]{0pt}{3.5ex}  Hard disk: size& yes  \\
\hline
\rule[-1ex]{0pt}{3.5ex}  Hard disk: IO speed & no  \\
\hline
\rule[-1ex]{0pt}{3.5ex}  Network: bandwidth & yes \\
\hline
\rule[-1ex]{0pt}{3.5ex}  Network: single stream speed & no(?) \\
\hline
\rule[-1ex]{0pt}{3.5ex}  CPU: cycles & yes  \\
\hline
\rule[-1ex]{0pt}{3.5ex}  CPU: single core speed  & no  \\
\hline
\rule[-1ex]{0pt}{3.5ex}  Databases & yes \\
\hline
\rule[-1ex]{0pt}{3.5ex}  Memory &  yes\\
\hline
\end{tabular}
\end{center}
\end{table}

In contrast to space missions which select the data-flow hardware upfront and stick with the choice, a project like ALMA, operating for 30 to 50  years will have to follow the general hardware evolution. Over such large timescales, vendor-lock-in is a serious risk. Scalability and flexibility in hardware and software solutions are thus mandatory. The ALMA hardware and the software stack have been chosen to be main-stream and off-the-shelf for that reason. The software stack is made out of free software as much as possible.

\subsection{EVOLUTION OF ASTRONOMY}
In full operations, ALMA will produce about the same amount of data in one year as ESO's telescopes have produced in its first 50 years. But ESO, too, will soon produce data at the same rate! Future facilities like LSST, with its predicted object catalogue of 20 billion entries with 200+ attributes each, LOFAR, SKA, PanStars, Euclid, Gaia, ELTs will all come with impressive data-rates. As Tony Tyson, LSST's Chief Scientist put it: "Astronomy is transformed from being a data-starved science to one where data is overabundant".

Large data-rates are per se not an issue, as computers for analysis will grow, too: The whole of the ALMA data will fit on a single standard consumer disk in already in mere 17 years from now in 2031. Also statistical analysis can be done on larger datasets in the exact same way as they can be on smaller ones. However, science which relies on visual inspection and individual analysis of spectra, images or data cubes will seriously be affected by the larger data-rates.

A precursor of this evolution could be seen in the Galaxy-Zoo project\footnote{http://www.galaxyzoo.org/} where visual inspection was impossible to do by the astronomers and was as a consequence outsourced to the public. However, a similar project with LSST galaxies instead of SDSS galaxies seems infeasible as essentially every literate person on the planet would have to help with the classification.

In addition to the larger data rates of individual observatories, in the recent years astronomy has already changed towards multi-wavelength science. A typical astronomer thus has "less time" for the analysis of data of a given wavelength range.

Whereas data will scale exponentially astronomers will not. Therefore the bytes per astronomer do scale exponentially. Our prediction is that whereas now astronomers are competing for observing time, in the future, observatories will be competing for astronomers. Observatories need the astronomers from the community in order to work with the observatories data and to publish results so that the observatories are successful and can secure further funding. Astronomers, not data, will be the rare resource.

Precursors of this evolution can already be observed. For example ALMA is meant to be a facility that can be used by astronomers without radio-astronomy background. In other words, it is a fundamental design principle of ALMA to "steal" astronomers from other communities.

\subsection{EVOLUTION OF SCIENCE}
As mentioned, we believe that astronomers will do less and less data-reduction and slowly also less of the initial analysis. This together with instruments becoming more complex and the general trend to more service-mode observing will transform astronomers into consumers instead of being co-producers of the data. On the one hand, this will allow scientists to concentrate much more on the science itself and will increase their productivity, and thus allows astronomy as a whole make the best use of astronomers, its future rare resource.

On the other hand, the risk that astronomers get too detached from the observing process and thus do not understand the technical limitations of the observations is real. Much of the responsibility for the quality of the results will be shifted from the astronomers to the observatories who will need to meticulously document the limitations of their data and workflow. With less instrument-specific knowledge in the community and more and more complex instruments, observatories will have to spend more effort training new staff to cover the observatories' needs for pipeline and post-analysis personnel.

Naturally, next to more automatized analysis using programmatic data and metadata access, statistical astronomy will become more prominent and data-mining techniques like machine-learning, clustering analysis and automatic classification will play a larger role.

The large amount of easily accessible science-grade data and the general overabundance likely make it easier for astronomers to pursue entire archival careers, i.e. to do astronomy without ever writing proposals.

We speculate that observatories will see their publication fractions (i.e. the fraction of proposals that been delivered to PIs, but did not get published) decrease and that they will start looking for means to make sure their investment gets indeed converted into science. This could be by having authors prove that they will have the resources available to analyze and publish the results, to have them justify why they did not publish if they did not, but -- above all -- to help them from the proposal preparation process, over the science-grade data products to archival research wherever possible. Given the relatively small effort it takes to submit a proposal asking for a lot of observing time, we believe that over-subscription factors are a relatively inaccurate indicator of the quality of a facility.

\section{SUMMARY}
\label{sec:summary}
ALMA, currently world's largest astronomy project, has started early science operations. At the centre of the ALMA data flow is the ALMA archive,  a database- and binary-data storage system distributed over three continents.

We argue that as astronomy is about to change from a data-starved science to a science where data is overabundant (T. Tyson), the way observatories operate changes. Driven by the exponential growth of astronomical data available per astronomer, observatories are and will be taking more and more of the workflow over from the astronomers, like data-reduction and a some of the science-analysis. We argue further that instead of being mere data-providers, they will (have to) modify the entire end-to-end user-experience for PIs (and archival researchers), from proposal-preparation over science-grade data-products to the science archive in order to stay successful in a world where observatories will compete for (the best) astronomers to work with their data -- and success is of course measured by the scientific output of the community. This will be golden times for astronomers.

The inventors of ALMA have carefully designed its operations model so that it fully follows the new paradigm: Extensive help is provided through the the ARCs, sophisticated tools make the submission of proposals as easy as possible and for the first time a radio-astronomy facility pledges to provide science-grade data products produced by an automatic pipeline.

Those data are searchable through a science archive that can be queried by physical concepts. In the future the Science Archive will have previews, VO-, reprocessing- and cutout services, a powerful server-side visualization as well as software providing an initial science-analysis of the data products.

\section{ACKNOWLEDGEMENT}
We gratefully acknowledge discussions with Daniel Durand, Alberto Micol, Jonas Haase and Martino Romaniello as well as the discussions with and the work for the ALMA archive of Andreas Wicenec, Gary Fuller, Sandra Etoka, Masao Saito, Holger Meuss, Robert Kurowski, Rodrigo Tobar, Karla Parussel, Paolo Nunes, Stefano Zampieri, Norm Hill, Viola Wang and Stewart Williams. ALMA is a partnership of ESO (representing its member states), NSF (USA) and NINS (Japan), together with NRC (Canada) and NSC and ASIAA (Taiwan), in cooperation with the Republic of Chile. The Joint ALMA Observatory is operated by ESO, AUI/NRAO and NAOJ.

\bibliography{spie_2014_stoehr_et_al}
\bibliographystyle{spiebib}

\end{document}